\documentclass[preprintnumbers,amsmath,amssymb,floatfix,9pt,prd,onecolumn,superscriptaddress,nofootinbib]{revtex4}
\usepackage{graphicx}
\usepackage{epsfig}
\usepackage{bm}
\usepackage{amsfonts}

\def\ben{\begin{equation}}
\def\een{\end{equation}}

 \def\bd{\begin{document}} \def\ed{\end{document}}
\def\ds{\documentstyle} \let\fr=\frac \let\bl=\bigl \let\br=\bigr
\let\Br=\Bigr \let\Bl=\Bigl
\let\bm=\bibitem
\let\na=\nabla
\let\pa=\partial \let\ov=\overline
\newcommand{\be}{\begin{equation}}
\newcommand{\ee}{\end{equation}}
\def\ba{\begin{array}}
\def\ea{\end{array}}
\def\ft#1#2{{\textstyle{\frac{\scriptstyle #1}{\scriptstyle #2} } }}
\def\fft#1#2{{\frac{#1}{#2}}}
\def\del{\partial}
\def\vp{\varphi}
\def\sst#1{{\scriptscriptstyle #1}}
\def\oneone{\rlap 1\mkern4mu{\rm l}}
\def\td{\tilde}
\def\wtd{\widetilde}
\def\ie{{\it i.e.\ }}
\def\dalemb#1#2{{\vbox{\hrule height .#2pt
        \hbox{\vrule width.#2pt height#1pt \kern#1pt
                \vrule width.#2pt}
        \hrule height.#2pt}}}
\def\square{\mathord{\dalemb{6.8}{7}\hbox{\hskip1pt}}}
\newcommand{\ho}[1]{$\, ^{#1}$}
\newcommand{\hoch}[1]{$\, ^{#1}$}
\newcommand{\bea}{\setlength\arraycolsep{2pt} \begin{eqnarray}}
\newcommand{\eea}{\end{eqnarray}}
\newcommand{\ra}{\rightarrow}
\newcommand{\lra}{\longrightarrow}
\newcommand{\Lra}{\Leftrightarrow}
\newcommand{\bp}{\tilde \beta^\prime}
\newcommand{\tr}{{\rm tr} }
\newcommand{\Tr}{{\rm Tr} }
\def\0{{\sst{(0)}}}
\def\1{{\sst{(1)}}}
\def\2{{\sst{(2)}}}
\def\3{{\sst{(3)}}}
\def\4{{\sst{(4)}}}
\def\5{{\sst{(5)}}}
\def\6{{\sst{(6)}}}
\def\7{{\sst{(7)}}}
\def\8{{\sst{(8)}}}
\def\m{{\sst{(m)}}}
\def\n{{\sst{(n)}}}
\def\cA{{{\cal A}}}
\def\cB{{{\cal B}}}
\def\cF{{{\cal F}}}
\def\cG{{{\cal G}}}
\def\cH{{{\cal H}}}
\def\tV{\widetilde V}
\def\tW{\widetilde W}
\def\tH{\widetilde H}
\def\tE{\widetilde E}
\def\tF{\widetilde F}
\def\tA{\widetilde A}
\def\im{{{\rm i}}}
\def\tY{{{\wtd Y}}}
\def\ep{{\epsilon}}
\def\vep{{\varepsilon}}
\def\bD{{{\bar D}}}
\def\R{{{\mathbb R}}}
\def\C{{{\mathbb C}}}
\def\H{{{\mathbb H}}}
\def\CP{{{\mathbb C}{\mathbb P}}}
\def\RP{{{\mathbb R}{\mathbb P}}}
\def\Z{{{\mathbb Z}}}
\def\bA{{{\mathbb A}}}
\def\bB{{{\mathbb B}}}
\def\bC{{{\mathbb C}}}
\def\bD{{{\mathbb D}}}
\def\bE{{{\mathbb E}}}
\def\bZ{{{\mathbb Z}}}
\def\Re{{{\frak{Re}}}}
\def\Im{{{\frak{Im}}}}
\def\cosec{{\,\hbox{cosec}\,}}
\def\Gm{{\Gamma_{\!\! -}}}
\def\Gp{{\Gamma_{\!\! +}}}
\def\stan{{standard }}
\def\nonstan{{supernumerary }}
\def\p{{\partial}}
\def\kdel#1{{\fft{\del}{\del#1}}}

\def\bog{{Bogomolny }}
\def\om{{\omega}}

\newcommand{\nnr}{\nonumber \\}
\newcommand{\pd}{\partial}
\newcommand{\ud}{\textrm{d}}
\newcommand{\dTH}{T^{\prime \, 0}_\textrm{H}}
\newcommand{\dOi}{\Omega^{\prime \, 0}_i}
\newcommand{\bx}{{\bf x}}
\begin{document}

\title{Cosmic Evolution in Brans-Dicke Chameleon Cosmology}
\author{\textbf{Mubasher Jamil}}
\email{mjamil@camp.nust.edu.pk} \affiliation{Center for Advanced
Mathematics and Physics (CAMP), National University of Sciences and
Technology (NUST), H-12, Islamabad, Pakistan}

\author{\textbf{Ibrar Hussain}}
\email{ibrar.hussain@seecs.nust.edu.pk} \affiliation{School of
Electrical Engineering and Computer Science (SEECS),\\
National University of Sciences and Technology (NUST), H-12,
Islamabad, Pakistan}

\author{\textbf{D. Momeni}}
\email{d.momeni@yahoo.com} \affiliation{ Department of Physics ,
Faculty of sciences, Tarbiat Moa'llem University, Tehran, Iran}

\begin{abstract} {\bf Abstract:} We have
investigated the Brans-Dicke Chameleon theory of gravity and
obtained exact solutions of the scale factor $a(t)$, scalar field
$\phi(t)$, an arbitrary function $f(\phi)$ which interact with the
matter Lagrangian in the action of the Brans-Dicke Chameleon theory
and potential $V(\phi)$ for different epochs of the cosmic
evolution. We plot the functions $a(t)$, $\phi(t)$, $f(t)$ and
$V(\phi)$ for different values of the Brans-Dicke parameter. In our
models, there is no accelerating solution, only decelerating one
with $q>0$. The physical cosmological distances have been
investigated carefully. Further the statefinder parameters pair and
deceleration parameter are discussed.
\end{abstract}

\maketitle

\section{Introduction}

The Brans-Dicke (BD) theory of gravity defined by a scalar field
$\phi$ and a constant coupling function $\omega$ \cite{BD1}, is
perhaps the most natural extension of general relativity (GR) which
is obtained in the limit of $\omega\rightarrow\infty$ and $\phi$=
constant \cite{KN,JDB,JMA}.  An imperative property of the BD theory
of gravity is that it yields simple expanding solutions \cite{CM}
for scalar field $\phi(t)$ and scale factor $a(t)$ which are
well-matched with solar system experiments \cite{PMG,SP,AGR}. There
are many works on interesting physical aspects of the BD
theory\cite{sergei}.

In a recent paper the dynamics of the cosmic evolution has been
investigated in the formalism of generalized BD theory of gravity
(where $\omega$ is not a constant but a function of $\phi(t)$)
\cite{singh}. There a consistent solution of the generalized BD
equations of motion based on simple power-law temporal behavior of
$a$, $\phi$ and $\omega$ is obtained.

In the present paper, a BD theory in which there is a non-minimal
coupling between the scalar field and the matter field is
considered. Thereby the action and the field equations are modified
due to the coupling of the scalar field with the matter. In the
literature such type of scalar field usually called "chameleon"
field \cite{JK}.We note that \emph{a chameleon field requires an
strict form of its potential in order to avoid the appearance of
fifth forces or violations of the Equivalence Principle, so that a
chameleon model has to present a realistic potential}.
 This is due to the fact that the physical
properties of the scalar field, such as its mass, depend delicately
on the environment. Moreover, in high density regions, the chameleon
mix together with its environment and becomes essentially invisible
to searches for Equivalence Principle violation and fifth force
\cite{JK}. Further more, it was shown that in the presence of
chameleon field, all existing constraints from planetary orbits,
such as those from lunar laser ranging are easily satisfied
\cite{JK,TP}. The explanation is that the chameleon-mediated force
between two large objects, such as the Earth and the Sun, is much
weaker than one would bluntly expect. In particular, it was shown
that the deviations from Newtonian gravity due to the chameleon
field of the Earth are suppressed by nine orders of magnitude by the
thin-shell effect \cite{TP}. Some other studies on the chameleon
gravity have been investigated in \cite{PB,SSG,KK}.

Our work differs from that of Ref. \cite{singh} in that we assume a
non-minimal coupling between the scalar field and the matter field.
besides, we are taking $\omega$ as a constant and not a function of
$\phi(t)$. Our aim is to investigate the dynamics of the potential
$V(\phi)$ and the function $f(\phi)$ (these are defined in the next
section) alongside the scale factor $a(t)$ and the scalar field
$\phi(t)$. In other words we are interested to find exact solutions
of $a(t)$, $\phi(t)$, $f(\phi)$ and $V(\phi)$ for different epochs
of the cosmic evolution. The stability analysis of the solutions is
important. But it will be beyond the scope of our paper. Such
analysis can be done using the similar methodology, as it has been
done in \cite{stability1}, by defining dimensionless variables. The
stability of BD with Chameleon field has been done already in
\cite{stability2} for the same system of equations as we have.

The statefinder parameters pair $\{r, s\}$, allows one to explore
the properties of dark energy (DE) independent of the model
\cite{sahni}. It has been used to distinguish flat models of the DE.
Recently this pair has been evaluated for different models
\cite{Zh1,We,Zh2,Hu,Zha,HuM,Zim,Sho,JD,SJ,Mde,Sharif}. In the
framework of Brans-Dicke Chameleon cosmology, the statefinder
parameters are studied in \cite{sr}.

Plan of the paper is as follows. In the next section we give the
basic equations of cosmic evolution with chameleon scalar field. In
the section III we discuss our model and obtain exact solution of
$a(t)$, $\phi(t)$, $f(\phi)$ and $V(\phi)$ for (A) radiation
dominated era, (B) dust fluid era and (C) vacuum energy dominated
era. In the section IV the statefinder parameters and the
deceleration parameter are investigated. The cosmological distances
have been discussed in section V. Finally we conclude our discussion
in section VI.

\section{Cosmic evolution with Chameleon scalar field\label{HDE}}
We begin with the  BD chameleon theory in which the scalar field is
coupled non-minimally to the matter field via the action
\cite{jamil} {
\begin{equation}
 S=\int{
d^{4}x\sqrt{-g}\left(\phi {R}-\frac{\omega}{\phi}g^{\mu
\nu}\partial_{\mu}\phi
\partial_{\nu}\phi-V(\phi) +f(\phi)L_m \right)},\label{act1}
\end{equation}}
where ${R}$ is the Ricci scalar curvature, $\phi$ is the BD scalar
field { with a potential $V(\phi)$. The chameleon field $\phi$ is}
non-minimally coupled to gravity, $\omega$ is the dimensionless BD
parameter. The last term in the action indicates the interaction
between the matter Lagrangian $L_m$ and some arbitrary function
$f(\phi)$ of the BD scalar field. In the limiting case $f(\phi) =1$,
we obtain the standard BD theory.

{ The gravitational field equations derived from the action
(\ref{act1}) with respect to the metric is
\begin{eqnarray}\label{feq}
R_{\mu\nu}-\frac{1}{2}g_{\mu\nu}R&=&\frac{f(\phi)}{\phi}T_{\mu\nu}+\frac{\omega}{\phi^2}\Big(\phi_\mu
\phi_\nu-\frac{1}{2}g_{\mu\nu}\phi^\alpha\phi_\alpha\Big)\nonumber\\&&
+\frac{1}{\phi}[\phi_{\mu;\nu}
-g_{\mu\nu}\Box\phi]-g_{\mu\nu}\frac{V(\phi)}{2\phi}.
\end{eqnarray}
where $T_{\mu\nu}$ represents the stress-energy tensor for the fluid
filling the spacetime which is represented by the perfect fluid
\begin{equation}\label{2aa}
T_{\mu\nu}=(\rho+p)u_\mu u_\nu+pg_{\mu\nu},
\end{equation}
where $\rho$ and $p$ are the energy density and pressure of the
perfect fluid which we assume to be a mixture of different kinds of
matters. Actually there is only chameleon field, whose pressure and
density are given by a perfect fluid stress tensor. For different
state parameters, this fluid behaves differently, like matter,
radiation or DE. Also $u^\mu$ is the four-vector velocity of the
fluid satisfying $u^\mu u_\mu=-1$. The Klein-Gordon equation (or the
wave equation) for the scalar field is
\begin{equation}\label{phi}
\Box\phi=\frac{T}{2\omega+3}\Big(f-\frac{1}{2}\phi
f_{,\phi}\Big)+\frac{1}{2\omega+3}( \phi V_{,\phi}-2V),
\end{equation}
where $T$ is the trace of (\ref{2aa})and $\Box=\nabla^\mu\nabla_\mu$
in which the operator $\nabla_\mu$ represents covariant derivative.
The homogeneous and isotropic Friedmann-Robertson-Walker (FRW)
universe is described by the metric
\begin{eqnarray}
 ds^2=-dt^2+a^2(t)\left(\frac{dr^2}{1-kr^2}+r^2d\Omega^2\right),\label{metric}
 \end{eqnarray}
where $a(t)$ is the scale factor, and  $k = -1, 0, +1$ corresponds
to open, flat, and closed universes, respectively. Variation of
action (\ref{act1}) with respect to metric (\ref{metric}) for a flat
$k=0$ universe filled with perfect fluid yields the following field
equations
\begin{eqnarray}
 &&H^2-\frac{\omega}{6}\frac{\dot{\phi} ^2}{\phi^2}+H
\frac{ \dot{\phi}}{\phi}=\frac{f(\phi)}{3\phi}\rho{+\frac{V(\phi)}{6\phi}},\label{FE1}\\
 &&2\frac{{\ddot{a}}}{a}+H^2+\frac{\omega}{2}\frac{\dot{\phi} ^2}{\phi^2}+2H
\frac{ \dot{\phi}}{\phi}+\frac{\ddot{\phi}}{\phi}
=-\frac{p}{\phi}{+\frac{V(\phi)}{2\phi}},\label{FE2}
\end{eqnarray}
where $H=\dot{a}/a$ is the Hubble parameter. Here, a dot indicates
differentiation with respect to the cosmic time $t$. The dynamical
equation (energy conservation) for the scalar field is{
\begin{equation}\label{FE3}
\ddot\phi+3H\dot\phi-\frac{\rho-3p}{2\omega+3}
\Big(f-\frac{1}{2}\phi f_{,\phi}\Big)+\frac{2}{2\omega+3}
\Big(V-\frac{1}{2}\phi V_{,\phi}\Big)=0.
\end{equation}}
Similarly the energy conservation for the cosmic fluid is
\begin{equation}\label{FE4}
\dot\rho+3H(\rho+p)=0.
\end{equation}
We shall use the equation of state (EoS) for the fluid
$p=\gamma\rho$, thus (\ref{FE4}) yields
\begin{equation}\label{FE40}
\rho=Ca^{-3(1+\gamma)}.
\end{equation}

\section{Our model}

Observational data of SN Ia suggests that theoretical models based
on power-law forms of the Chameleon potential $V(\phi)$ and scalar
function $f(\phi)$ are consistent with the data \cite{sr}. Hence we
shall follow the procedure of \cite{singh} and will obtain solution
of the above dynamical equations (\ref{FE1}) to (\ref{FE4}), by
assuming power law dependence on time for $a(t)$, $\phi(t)$,
$f(\phi)$ and $V(\phi)$.
\begin{equation}\label{FE5}
a(t)=a_0\Big(\frac{t}{t_0}\Big)^\alpha,
\end{equation}
\begin{equation}\label{FE6}
\phi(t)=\phi_0\Big(\frac{t}{t_0}\Big)^\beta,
\end{equation}
\begin{equation}\label{FE7}
f(\phi(t))\sim \phi(t)^n=f_0\Big(\frac{t}{t_0}\Big)^{n\beta},
\end{equation}
\begin{equation}\label{FE8}
V(\phi(t))\sim \phi(t)^m=V_0\Big(\frac{t}{t_0}\Big)^{m\beta}.
\end{equation}
Note that $\omega$ is a constant BD parameter and $a_0$, $\phi_0$,
$f_0$ and $V_0$ are also constants. Also notice that dynamical
system is a closed system i.e. four differential equations
(\ref{FE1}) to (\ref{FE4}) for four unknown parameters ($\alpha$,
$\beta$, $m$, $n$) to be determined. $\alpha$ must be positive for
an expanding Universe while other parameters are free. Using
(\ref{FE5}) in (\ref{FE40}), we get
\begin{equation}\label{FE41}
\rho=\rho_0\Big(\frac{t}{t_0}\Big)^{-3\alpha(1+\gamma)},\ \ \
p=\gamma\rho_0\Big(\frac{t}{t_0}\Big)^{-3\alpha(1+\gamma)}.
\end{equation}
Using the set of ansatz functions(11-15) we get,
\begin{eqnarray}
\alpha^2-\frac{\omega}{6}\beta^2+\alpha\beta=0,\\
f_0\rho_0=-\frac{V_0}{2},\\
(n-m)\beta=3\alpha(1+\gamma).
\end{eqnarray}

We now proceed to check the consistency of above equations with
Eq.(8) which is the wave equation for the scalar field $\phi(t)$.
Using eqs. (11-15)    Eq.(8) reduces to
\begin{eqnarray}
\beta(\beta-1+3\alpha)=0,\\
(2-m)V_0=\rho_0 f_0 (2-n)(1-3\gamma),\\
(m-n)\beta=-3\alpha(1+\gamma).
\end{eqnarray}
The above Eq. (19) implies that we can have $\beta=0$ or
$\beta=1-3\alpha$. The first one results that
\begin{eqnarray}
\phi(t)=\phi_0,
\\f(\phi)=f_0,\\
V(\phi)=V_0.
\end{eqnarray}
The above set of functions or their equivalent value $\beta=0$ in
(16) gives us $\alpha=0$. But we remove $\beta=0$ in our toy model
we must take $\alpha>0$. Thus in the present work we discard this
special case and limit ourselves only to
\begin{eqnarray}
\beta=1-3\alpha
\end{eqnarray}
Now we analyze different cosmological epochs. That is the cases of
$\gamma=\frac{1}{3}$, $\gamma=0$. We remove completely the case with
$\gamma=-1$, since as we pointed it previously , in our models,
there is no accelerating solution, only decelerating one with $q>0$.

\subsection{Radiation dominated ($\gamma=1/3$)}
In this case Eq. (20)gives $m=2$. Substituting $\beta=1-3\alpha$ in
(16) we obtain
\begin{eqnarray}
\alpha=\frac{3(\omega+1)\pm\sqrt{6\omega+9}}{12+9\omega},\\
\beta=\frac{1\mp\sqrt{6\omega+9}}{4+3\omega}.
\end{eqnarray}
Now from (18) we obtain
\begin{eqnarray}
n=\frac{14+12\omega\pm2\sqrt{6\omega+9}}{1\mp\sqrt{6\omega+9}}
\end{eqnarray}
Thus we have
\begin{equation}
a(t)=a_0\Big(\frac{t}{t_0}\Big)^{\frac{3(\omega+1)\pm\sqrt{6\omega+9}}{12+9\omega}},
\end{equation}
\begin{equation}
\phi(t)=\phi_0\Big(\frac{t}{t_0}\Big)^{\frac{1\mp\sqrt{6\omega+9}}{4+3\omega}},
\end{equation}
\begin{equation}
f(\phi(t))=f_0\Big(\frac{t}{t_0}\Big)^{\frac{14+12\omega\pm2\sqrt{6\omega+9}}{4+3\omega}},
\end{equation}
\begin{equation}
V(\phi(t))=V_0\Big(\frac{t}{t_0}\Big)^{2\frac{1\mp\sqrt{6\omega+9}}{4+3\omega}}.
\end{equation}
Below we plot some figures include the time behavior of the set of
functions $a(t)$, $\phi(t)$, $f(\phi(t))$, $V(\phi(t))$ for some
large values of the parameter $\omega$. The first two figures show
the accelerated expansion of Universe for some large values of the
BD parameter.
\begin{figure}
\centering
\includegraphics[scale=0.4]{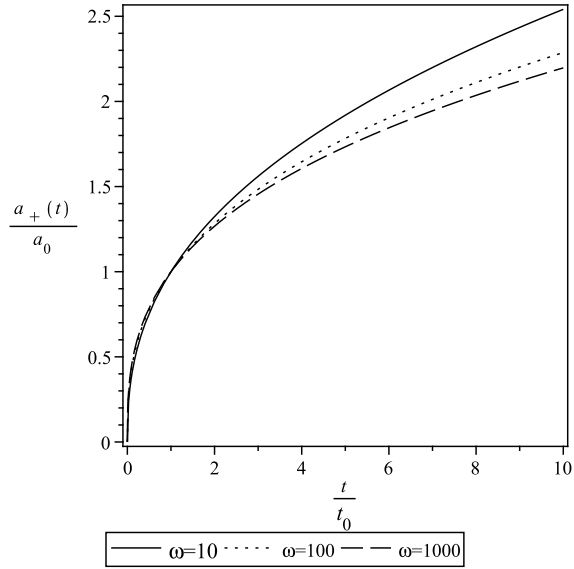} % scale goes from 0 to 1.
\caption{ Variation of scale factor  $\frac{a_{+}(t)}{a_{0}}$ for
different values of the BD parameter $\omega$.} \label{fig2.eps}
\end{figure}

\begin{figure}
\centering
 \includegraphics[scale=0.4]{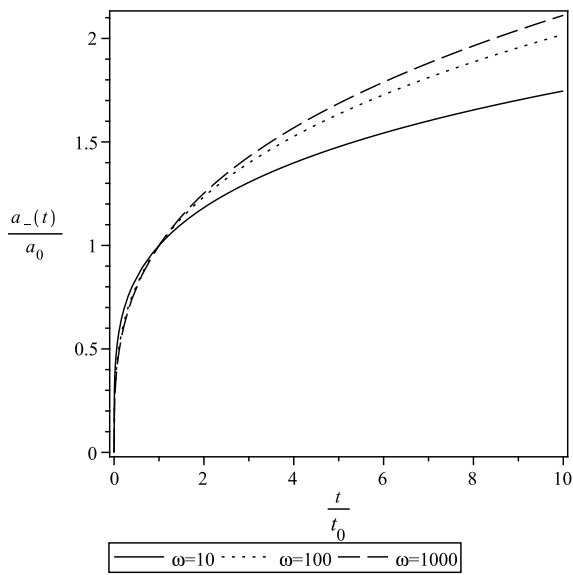} % scale goes from 0 to 1.
 \caption{ Variation of scale factor  $\frac{a_{-}(t)}{a_{0}}$ for different values of the BD parameter $\omega$.}
  \label{fig3.eps}
\end{figure}
The next two figure show the two possible values of the scalar field
for a sample of large BD parameter.
\begin{figure}
\centering
 \includegraphics[scale=0.4]{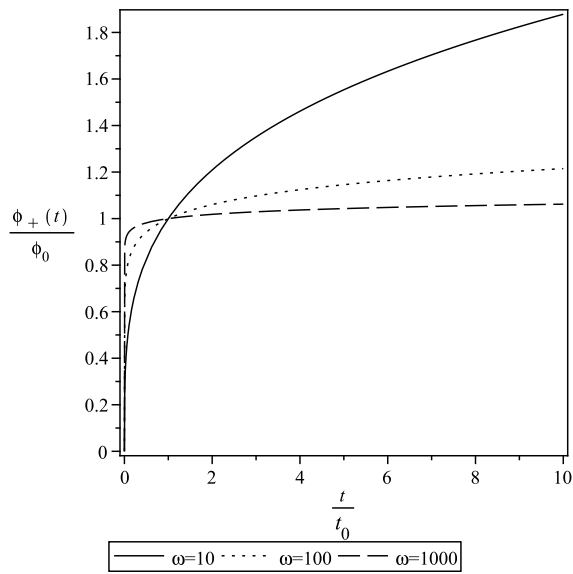} % scale goes from 0 to 1.
 \caption{ Variation of  the one possible value of the scalar field
for a sample of large BD parameter $\omega$.}
  \label{fig4.eps}
\end{figure}

\begin{figure}
\centering
 \includegraphics[scale=0.4]{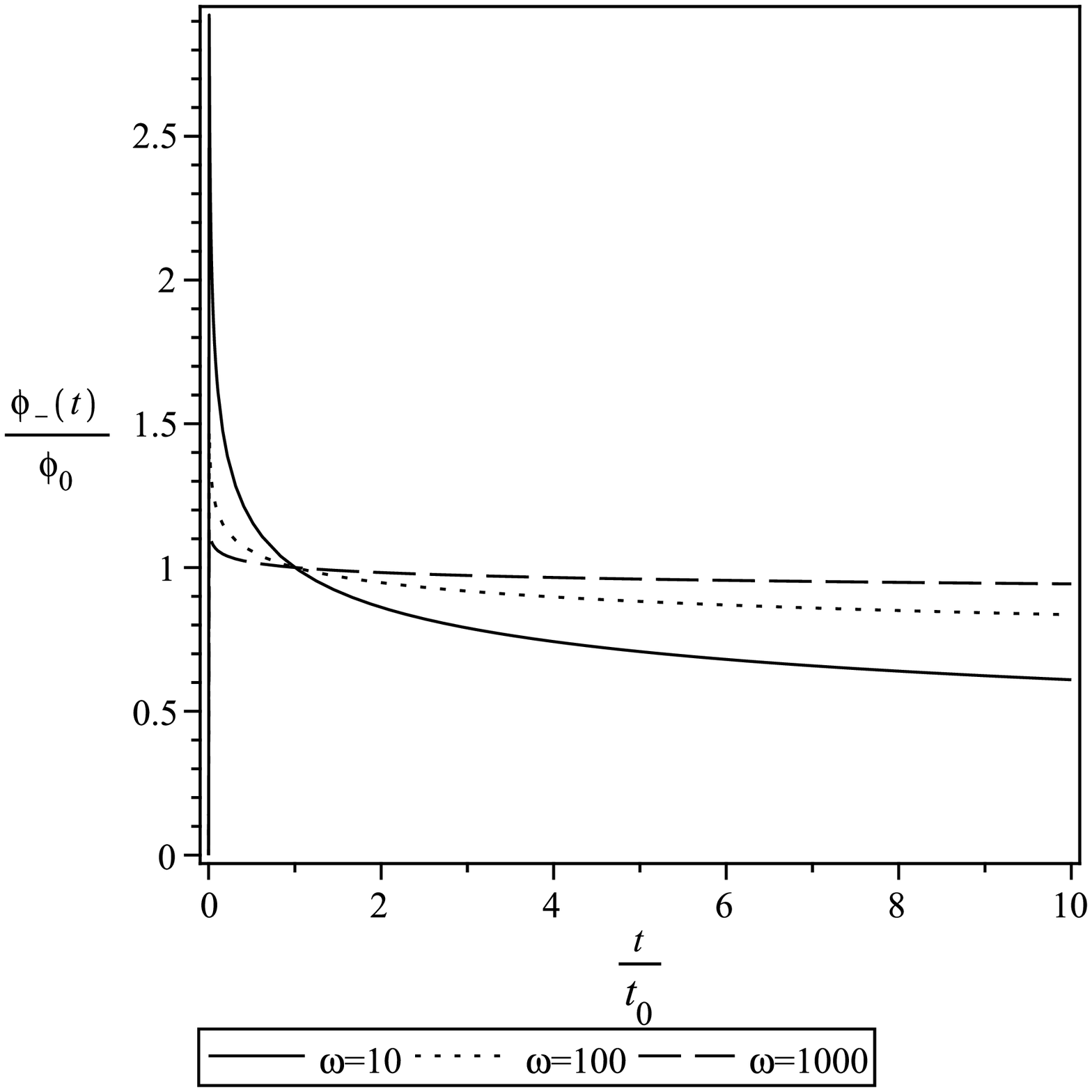} % scale goes from 0 to 1.
 \caption{ Variation of  the another  possible value of the scalar field
for a sample of large BD parameter $\omega$.}
  \label{fig5.eps}
\end{figure}

\begin{figure}
\centering
 \includegraphics[scale=0.4]{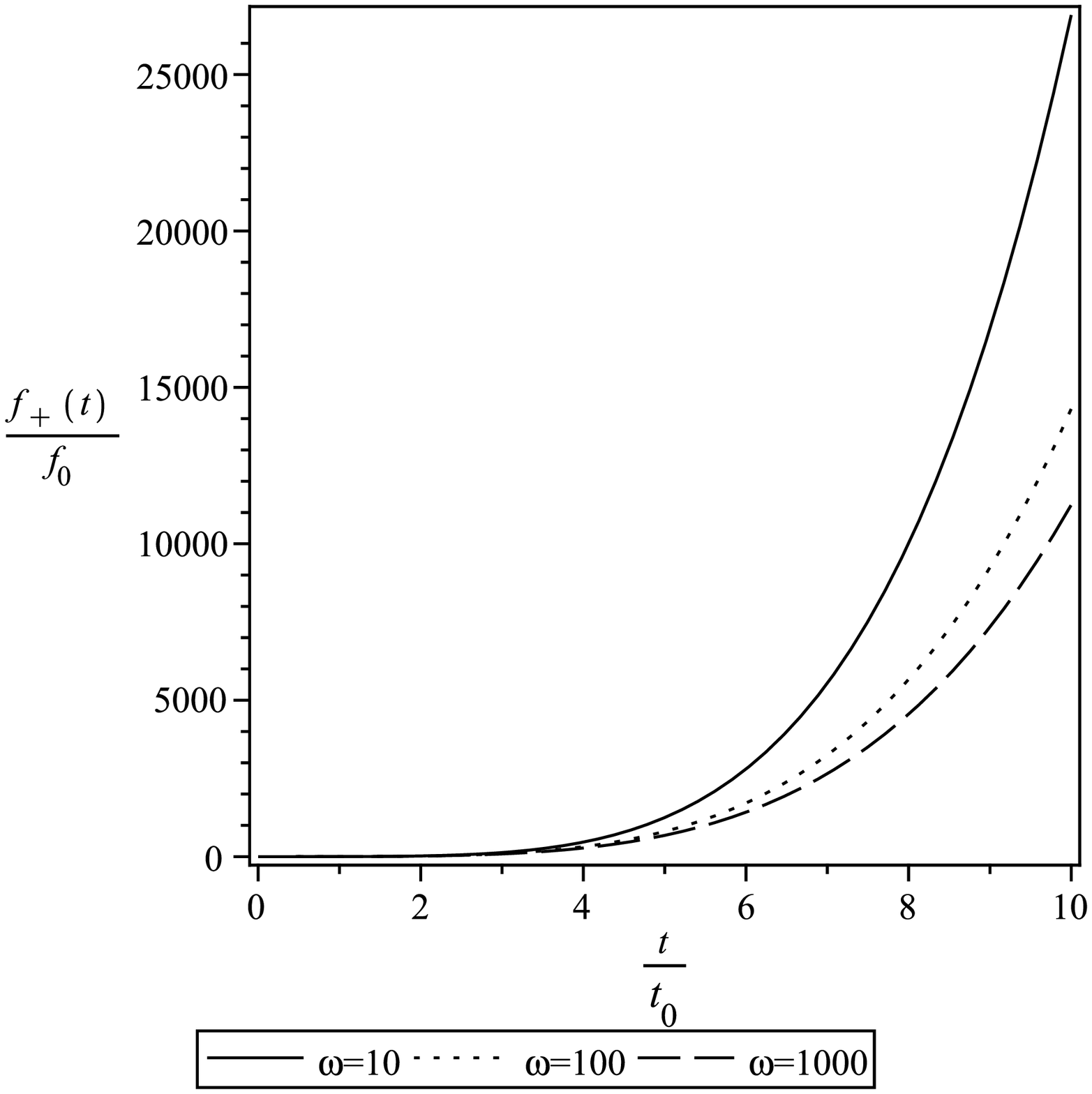} % scale goes from 0 to 1.
 \caption{ Variation of  $f_{+}(t)$
for a sample of large BD parameter $\omega$.}
  \label{fig6.eps}
\end{figure}

\begin{figure}
\centering
 \includegraphics[scale=0.4]{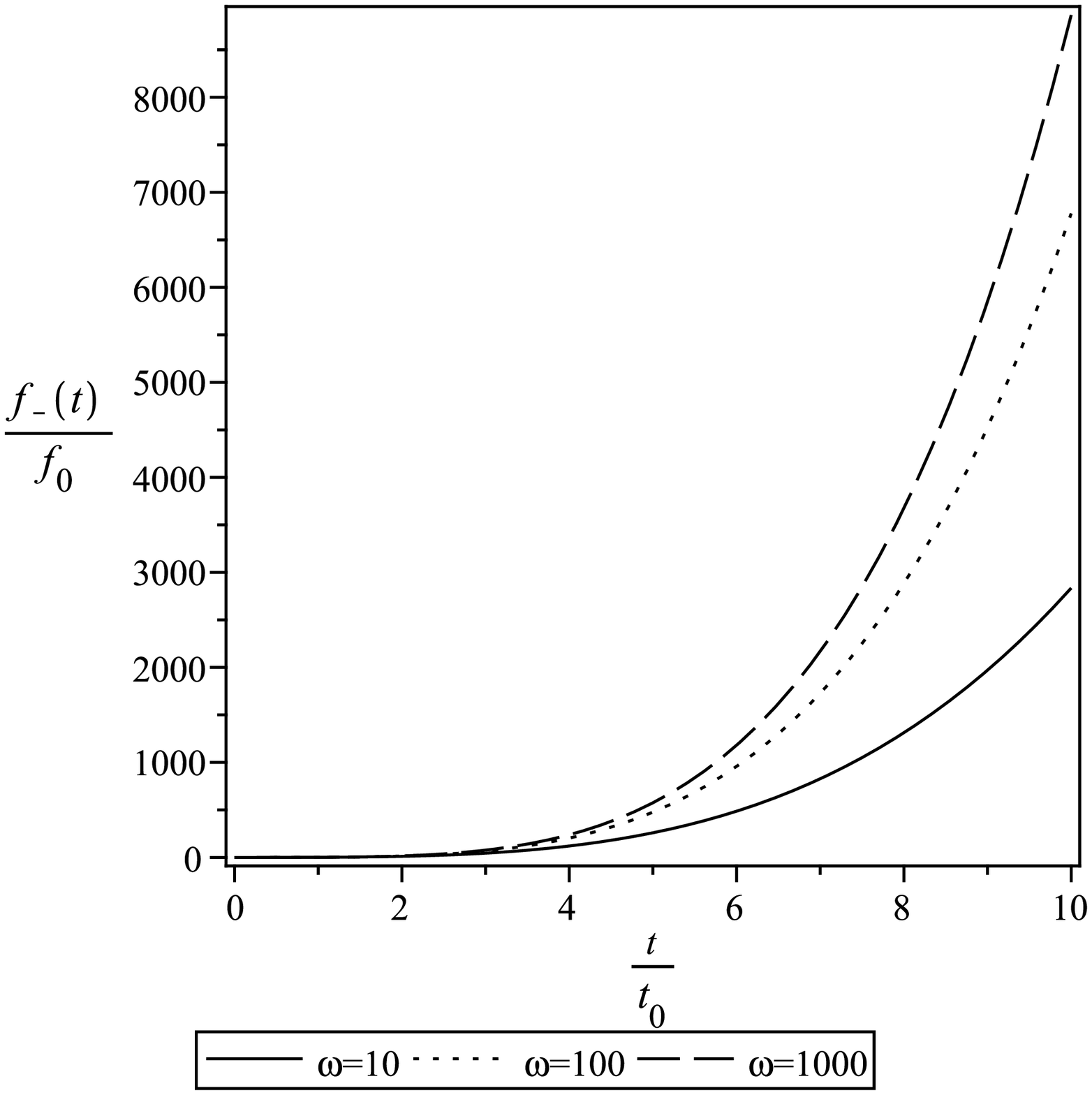} % scale goes from 0 to 1.
 \caption{ Variation of  $f_{-}(t)$
for a sample of large BD parameter $\omega$.}
  \label{fig7.eps}
\end{figure}

\begin{figure}
\centering
 \includegraphics[scale=0.4]{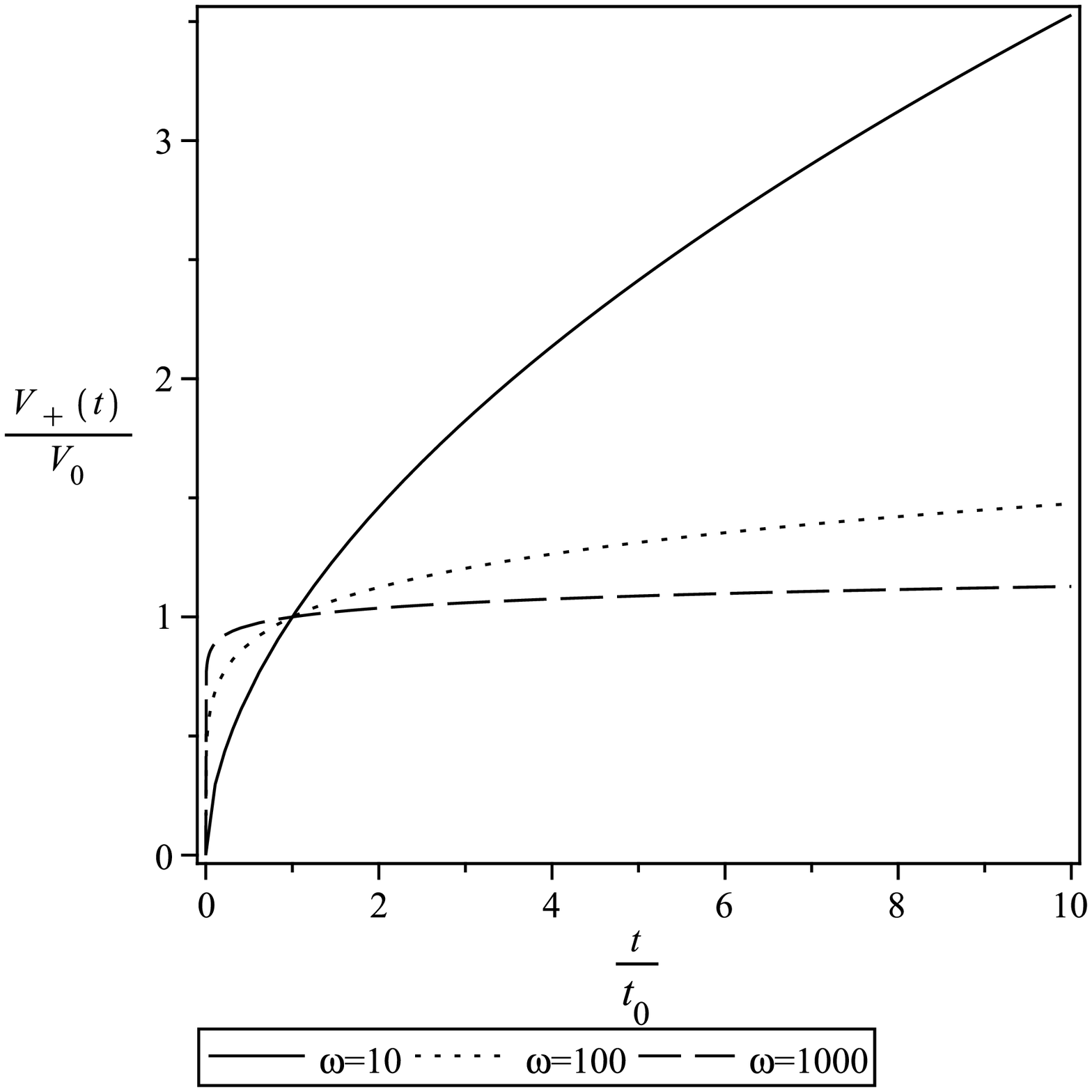} % scale goes from 0 to 1.
 \caption{ Variation of  $V_{+}(t)$
for a sample of large BD parameter $\omega$.}
  \label{fig8.eps}
\end{figure}

\begin{figure}
\centering
 \includegraphics[scale=0.4]{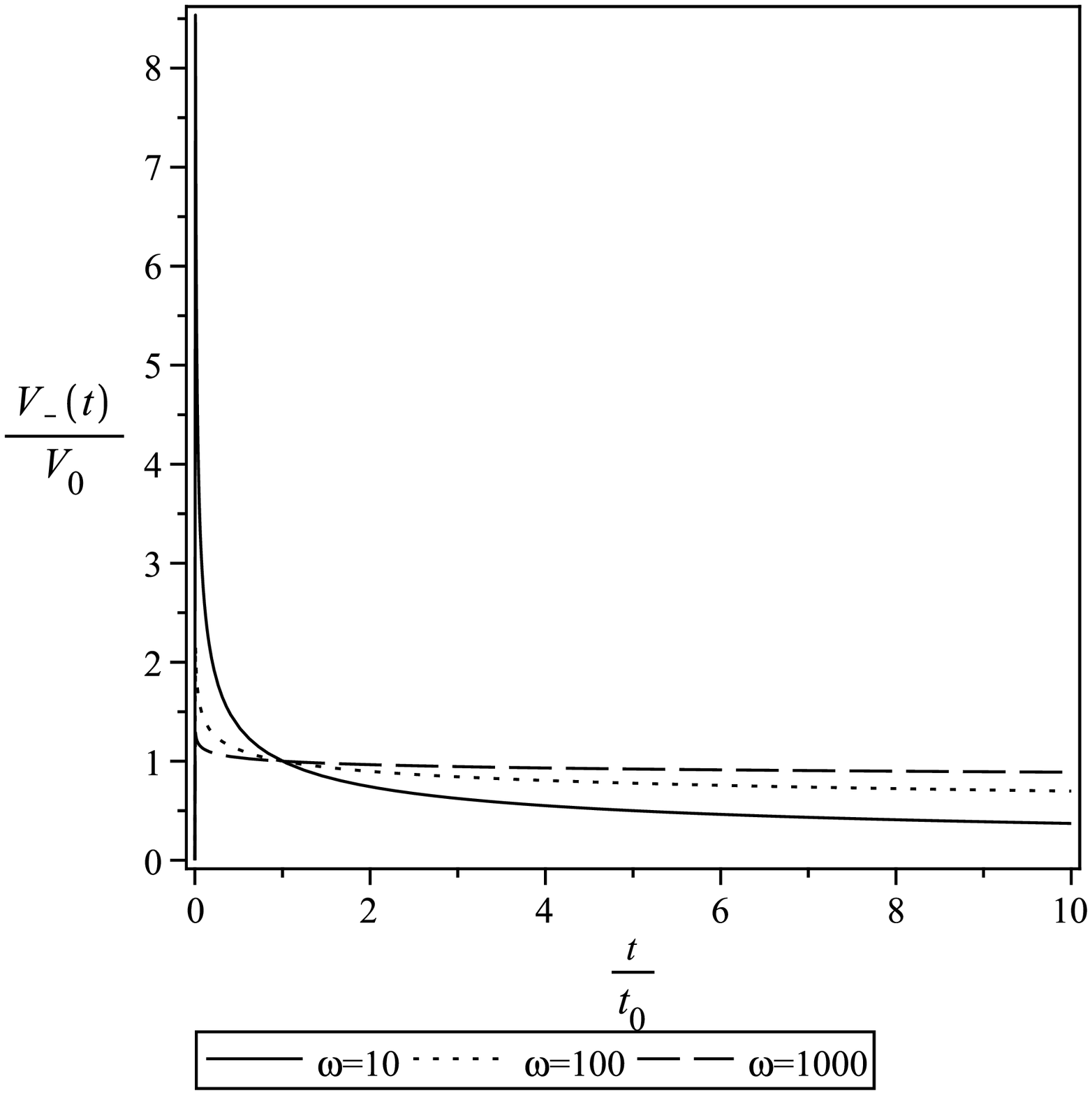} % scale goes from 0 to 1.
 \caption{ Variation of  $V_{-}(t)$
for a sample of large BD parameter $\omega$.}
  \label{fig9.eps}
\end{figure}
 We obtain that the radiation dominant toy model is a simple
 square power model. Here we plot the potential function as a
 function of the scalar field for some values of the $\phi_0,V_0$.
\begin{figure}
\centering
 \includegraphics[scale=0.4]{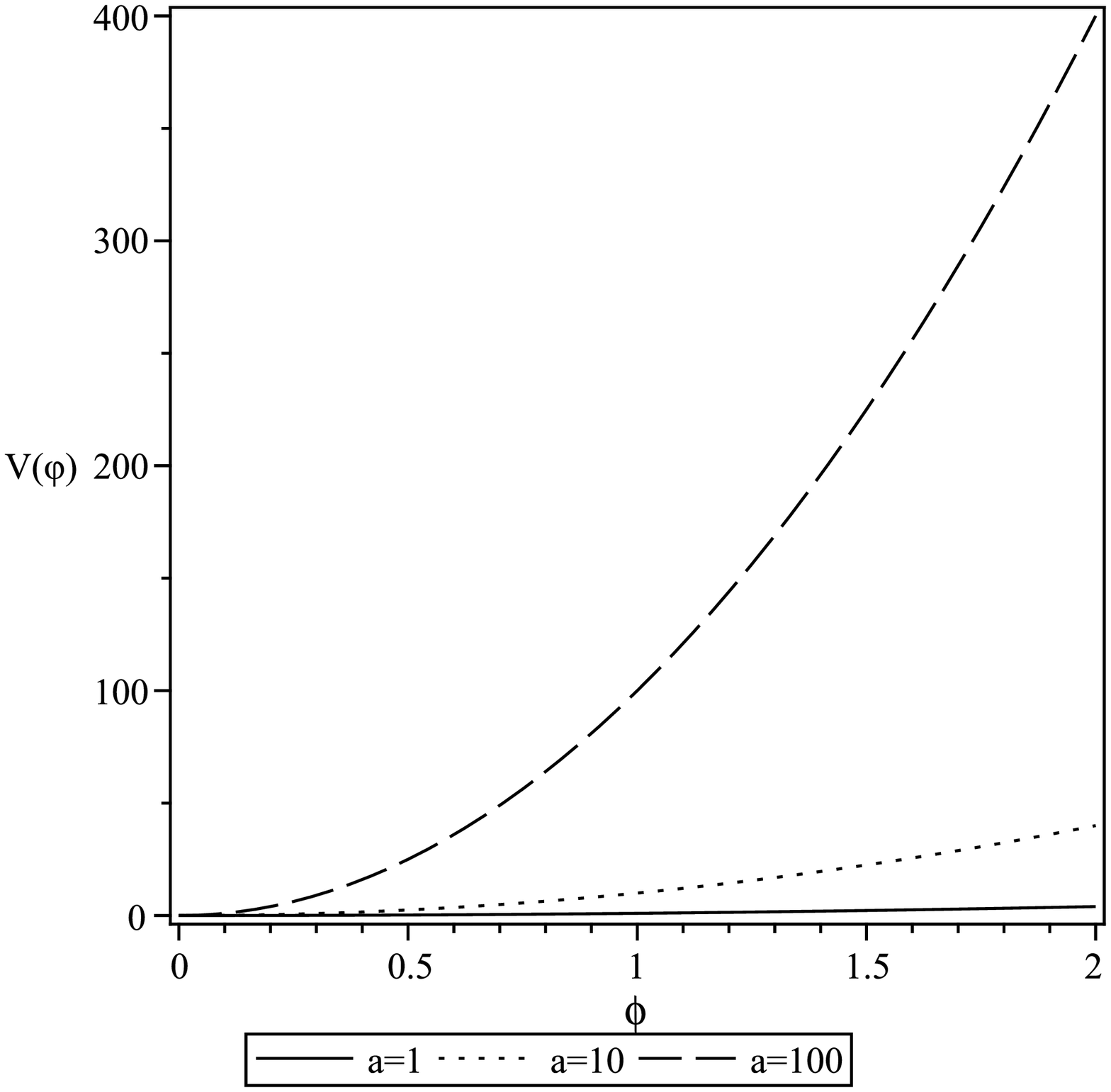} % scale goes from 0 to 1.
 \caption{ Variation of  $V(\phi)$
for a sample of the parameter $a=\frac{V_0}{(\phi_0)^2}$.}
  \label{fig10.eps}
\end{figure}

Remind that for the limit of the very large BD parameter $\omega$
which we expect that the theory must has a GR limit , the exponents
and the functions are
\begin{eqnarray}
\alpha_{GR}(t)=\frac{1}{3}\\
\beta\simeq 0,\\
n\simeq 0.
\end{eqnarray}

\begin{eqnarray}
a_{GR}(t)=a_0\Big(\frac{t}{t_0}\Big)^{\frac{1}{3}}
\end{eqnarray}
\begin{equation}
\phi(t)\simeq\phi_0,
\end{equation}
\begin{equation}
f(\phi(t))\simeq f_0,
\end{equation}
\begin{equation}
V(\phi(t))\simeq V_0.
\end{equation}
The plot of the scale factor for the GR limit is Fig.(10)
\begin{figure}
\centering
 \includegraphics[scale=0.4]{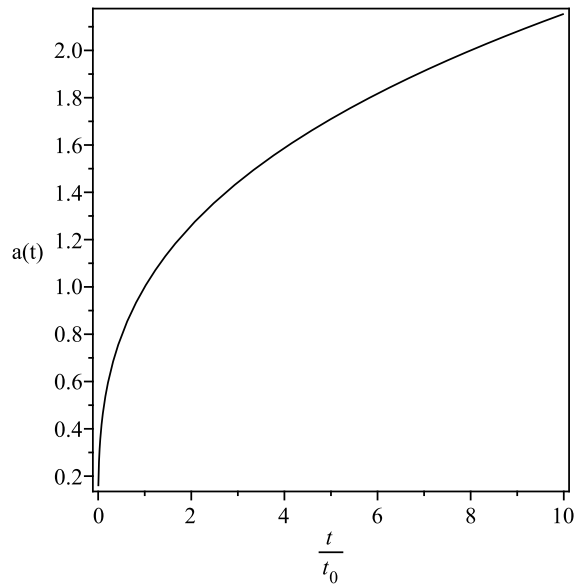} % scale goes from 0 to 1.
 \caption{ Variation of scale factor in GR limit}
  \label{fig1.eps}
\end{figure}

%\begin{figure}
%\centering
 %\includegraphics[scale=0.4]{fig4} % scale goes from 0 to 1.
  %\caption{ The statefinder parameters for some values of $n$.}
  %\label{fig4.eps}
%\end{figure}

\subsection{ Dust fluid ($\gamma=0$)}
This case from (20,21) we've
\begin{eqnarray}
m+\frac{n}{2}=3\\
(m-n)=-\frac{9(\omega+1)\pm3\sqrt{6\omega+9}} {1\mp\sqrt{6\omega+9}}
\end{eqnarray}
from (40,41) we obtain
\begin{eqnarray}
n=\frac{2(3\omega+4)}{1\mp\sqrt{6\omega+9}}\\
m=\frac{-(1+3\omega)\mp3\sqrt{6\omega+9}}{1\mp\sqrt{6\omega+9}}
\end{eqnarray}
Thus we have
\begin{eqnarray}
f(\phi(t))=f_0\Big(\frac{t}{t_0}\Big)^{2}
\end{eqnarray}

\begin{eqnarray}
V(\phi(t))=V_0\Big(\frac{t}{t_0}\Big)^{\frac{-(1+3\omega)\mp3\sqrt{6\omega+9}}
{4+3\omega}}.
\end{eqnarray}
for large values of the BD parameter $\gamma$, we have
$V(\phi(t))=V_0\Big(\frac{t}{t_0}\Big)^{-1}$.

\section{Statefinder parameters and deceleration parameter}

The statefinder parameters $r$ and $s$ depends on the third and
second derivatives of the scale factor $a$, just as the dependence
of the Hubble parameter $H$ and the deceleration parameter $q$ on
its first and second derivatives respectively.

The deceleration parameter is defined as
\begin{equation}
q=-\frac{\ddot a}{a\dot a^2}.
\end{equation}
The statefinder parameters are \cite{sahni}
\begin{equation}
r=\frac{\dddot a a^2}{\dot a^3},\ \ s=\frac{r-1}{3(q-\frac{1}{2})}.
\end{equation}
For Eq. (29), the above three parameters take the form
\begin{eqnarray}
q_\pm&=&\frac{3+2\omega\mp\sqrt{9+6\omega}}{\omega},\nonumber \\
r_\pm&=&\frac{(\mp21\mp15\omega+\sqrt{9+6\omega})(\mp9\mp6\omega+\sqrt{9+6\omega})}
{(3+3\omega+\sqrt{9+6\omega})^2},\nonumber \\
s_\pm&=& \frac{6+6\omega\mp\sqrt{9+6\omega}}{3\omega}.
\end{eqnarray}

\begin{figure}
\centering
 \includegraphics[scale=0.4]{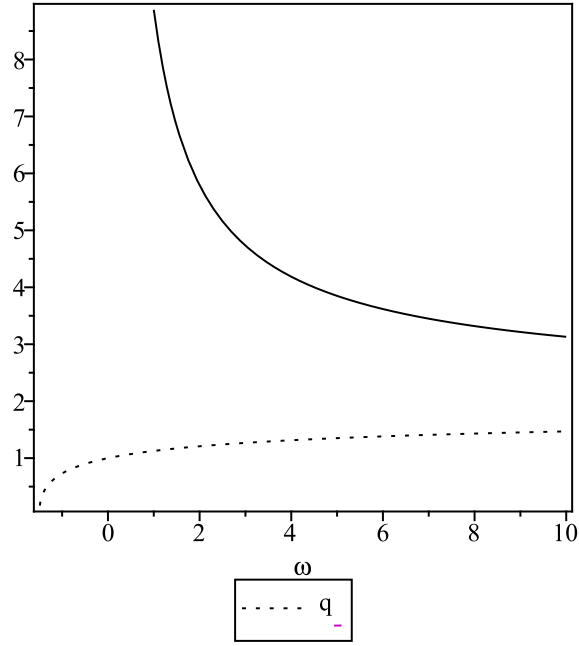} % scale goes from 0 to 1.
 \caption{ Variation of the deceleration parameter $q_{+,-}$. The line is
 $q_{+}$ and dot denotes  $q_{-}$.}
  \label{fig11.eps}
\end{figure}

\begin{figure}
\centering
 \includegraphics[scale=0.4]{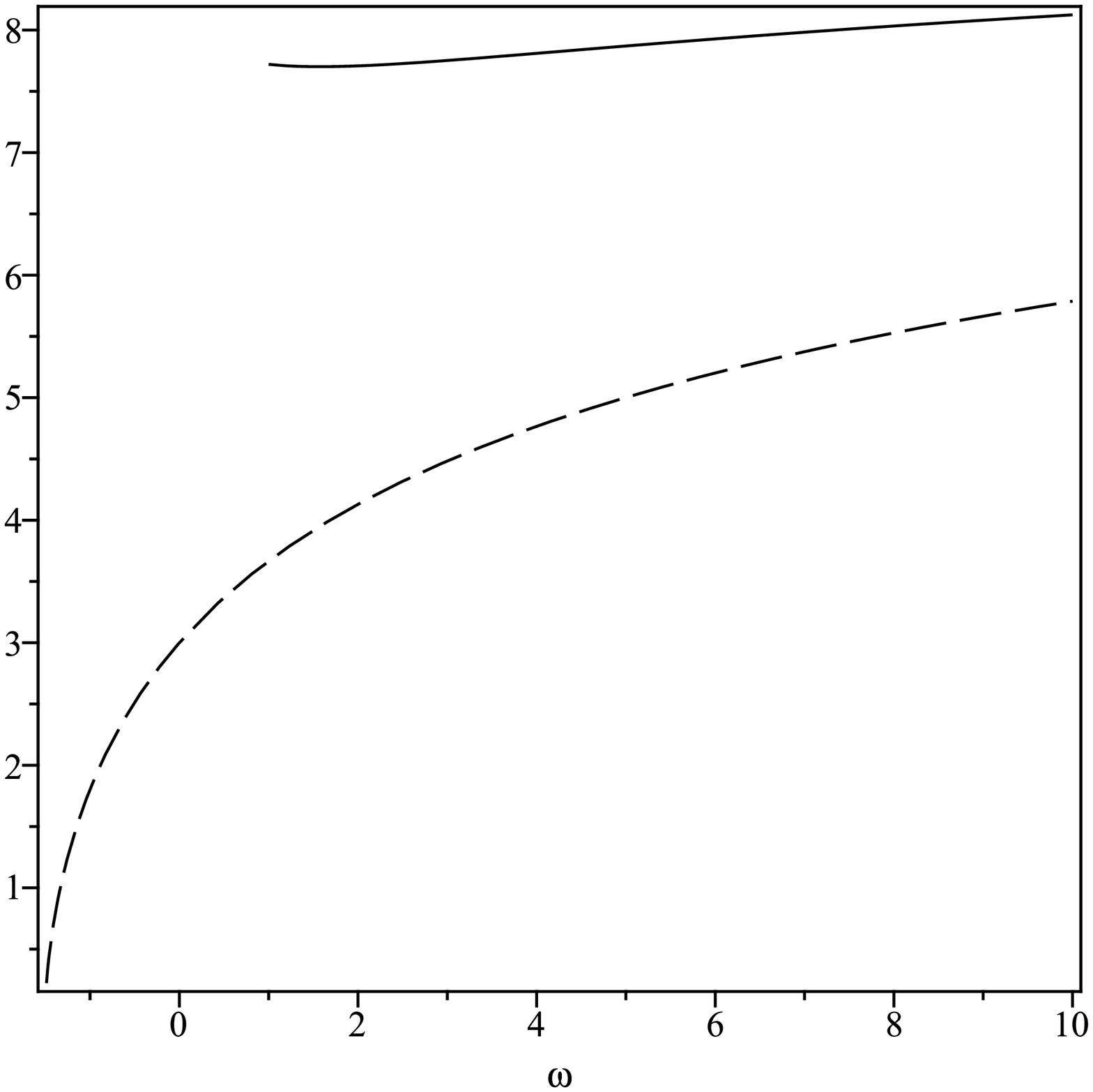} % scale goes from 0 to 1.
 \caption{ Variation of the  parameter $r_{+,-}$. The line is
 $r_{+}$ and dash denotes  $r_{-}$.}
  \label{fig12.eps}
\end{figure}

\begin{figure}
\centering
 \includegraphics[scale=0.4]{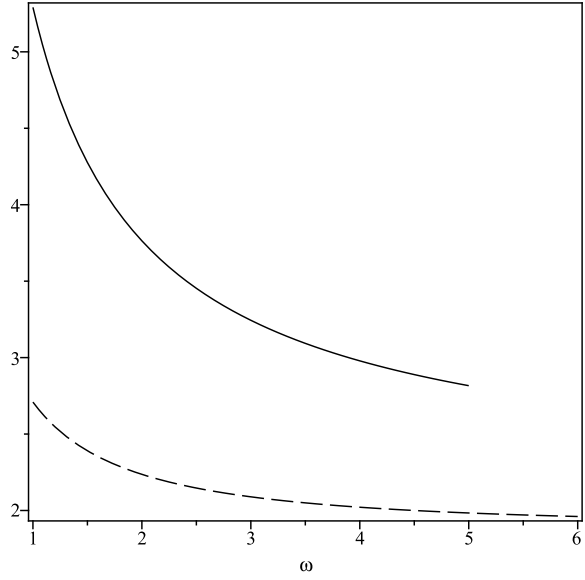} % scale goes from 0 to 1.
 \caption{ Variation of the  parameter $s_{+,-}$. The line is
 $s_{+}$ and dash denotes  $s_{-}$.}
  \label{fig13.eps}
\end{figure}

\section{Cosmological distances}
In this section we discuss the time dependent cosmological distances
of the models presented in sections A,B.

\subsection{Lookback time}
 If a photon is emitted by a source at the
instant $t$ and received at time $t_0$, then the photon travel time
or the lookback time $t- t_0$ is defined by
%%%
\begin{eqnarray}
 t-t_0=-\int_{0}^{z}\frac{dz}{H(z)
(1+z)},
\end{eqnarray} %%%

where $a_0$ is the present value of the scale factor of the
Universe. If a photon emitted by a source and received by an
observer at time $t_0$ then the proper distance between them is
defined by
%%%
\begin{eqnarray}
 t-t_0=(1+z)^{-\frac{1}{\alpha}}.
   \end{eqnarray}

 The Figure-14 shows the behavior of the lookback time as a
function of the redshift for value different values of the $\omega$
and for the plus sign in (26).

\begin{figure}
\centering
 \includegraphics[scale=0.5]{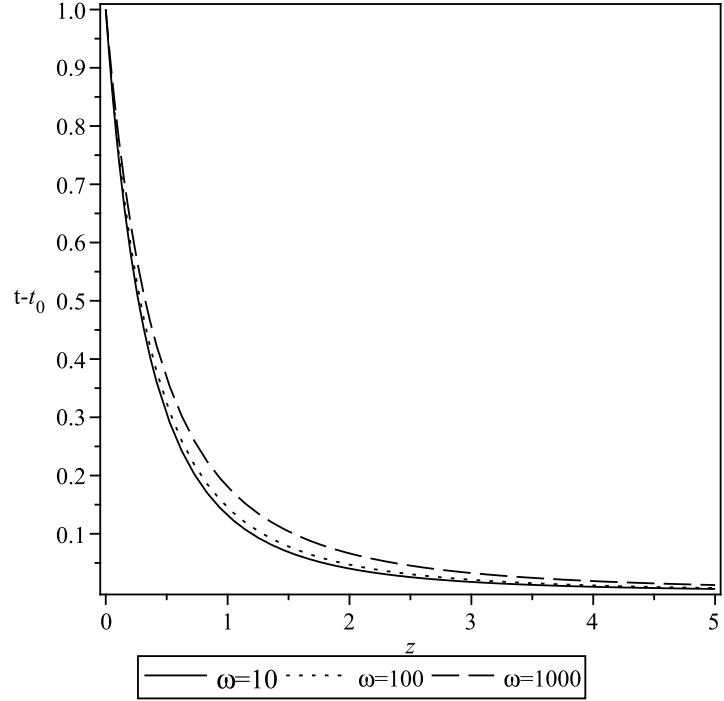} % scale goes from 0 to 1.
  \caption{ The  behavior of the lookback time as a
function of the redshift for values $\omega$.}
 \label{fig14.eps}
\end{figure}

\subsection{Proper distance}
If a photon emitted by a source and received by an observer at time
$t_0$ then the proper distance between them is defined by
%%%
\begin{eqnarray}
 d=a_0\int_{a}^{a_0}\frac{da}{aH^2}=\int_{0}^{z}\frac{dz}{H^2(z)(1+z)},
 \end{eqnarray}
%%%1
which for (29) simplifies to
%%%
\begin{eqnarray}
d=\frac{1}{2\alpha}(1-\frac{1}{(1+z)^{\frac{2}{\alpha}}}).
    \end{eqnarray} %%%

\begin{figure}
\centering
 \includegraphics[scale=0.5]{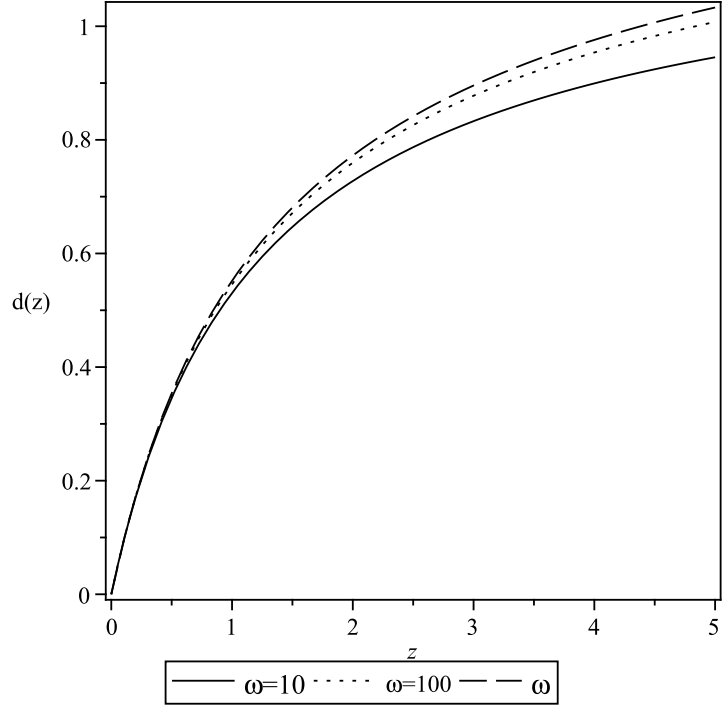} % scale goes from 0 to 1.
  \caption{ The  behavior of the proper distance as a
function of the redshift for values $\omega$.}
  \label{fig15.eps}
\end{figure}

\subsection{Luminosity distance}
If $L$ be the total energy emitted by the source per unit time and
$\ell$ be the apparent luminosity of the object then the luminosity
distance $d_{L}=\left(\frac{L}{4\pi\ell}\right)^{\frac{1}{2}}$
evolves as
%%%
\begin{eqnarray}
d_{L}=d(1+z).
\end{eqnarray} %%%

The figure (16) shows the variation of
 the luminosity distance as a function of the exponent $\omega$.

\begin{figure}
\centering
 \includegraphics[scale=0.5]{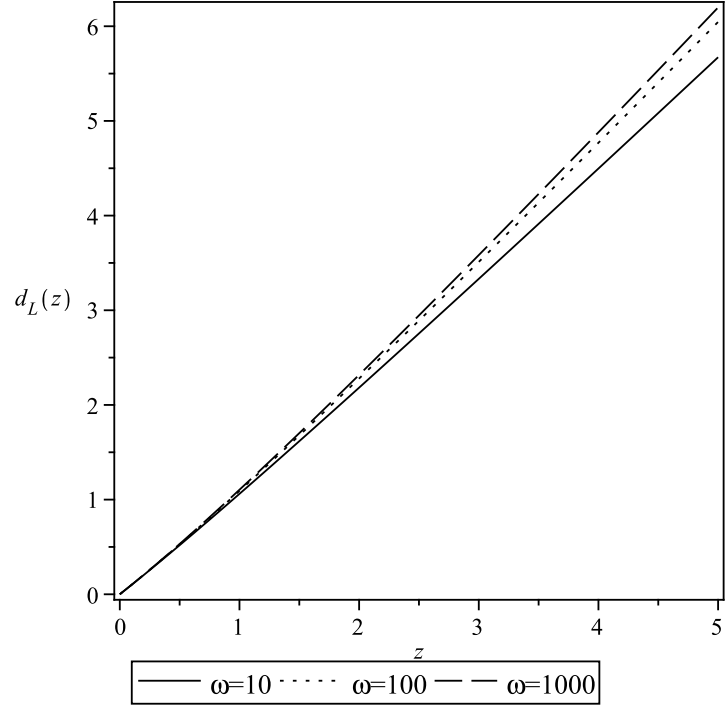} % scale goes from 0 to 1.
  \caption{ The  behavior of the luminosity distance as a
function of the redshift for values $\omega$.}
  \label{fig16.eps}
\end{figure}

\subsection{Angular diameter}
The angular diameter distance $d_A$ is simplified to
%%%
\begin{eqnarray} d_A=d_L
(1+z)^{-2}.
\end{eqnarray}
 %%%

The Figure-17 shows the variation of the angular diameter as a
function of the exponent $0<n<1$.
\begin{figure}
\centering
 \includegraphics[scale=0.5]{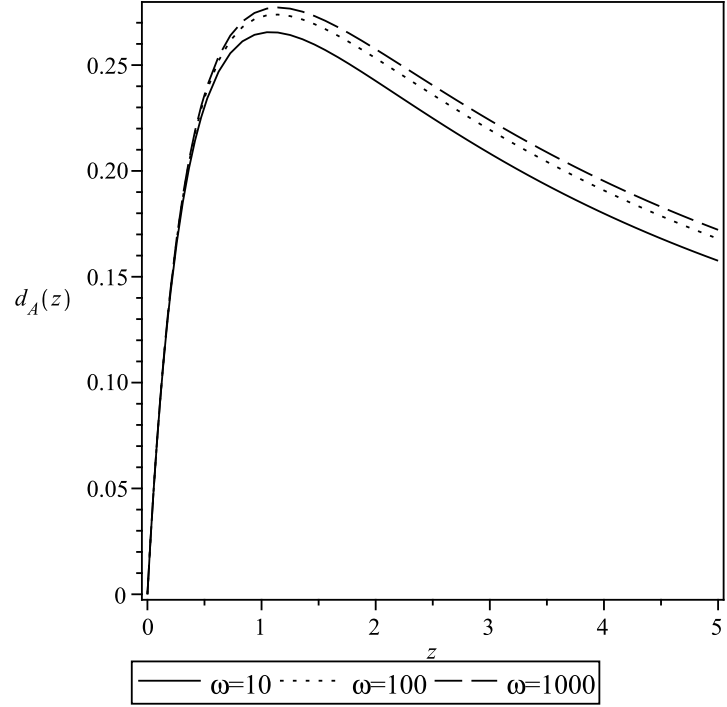} % scale goes from 0 to 1.
  \caption{ The  behavior of the angular diameter as a
function of the redshift for values $\omega$.}
  \label{fig17.eps}
\end{figure}

\section{Conclusion}

To recapitulate, we studied the Brans-Dicke Chameleon cosmology by
obtaining exact solutions of the scale factor $a(t)$, scalar field
$\phi(t)$, the potential $V(\phi)$ and the arbitrary function
$f(\phi)$ for different epochs of the cosmic evolution. This
analysis was performed by using ansatz for the above parameters.
These were motivated since earlier studies of dark energy show that
these cosmic parameters obey power-law form of the time parameter.
Next we plotted these parameters for various values of the BD
parameter in different cosmic epochs.

In figures 1 and 2, the scale factor is plotted is shown for
different times. The positive and negative subscripts of $a(t)$
correspond to two values given in (29). The behavior of scalar field
$\phi$ is plotted in figures 3 and 4. It is shown that the positive
component of the field $\phi_+$ increases faster for small values of
$\omega$ and vice versa for $\phi_-$. In figures 5 and 6, the
expressions for $f_+$ and $f_-$ are plotted. It is observed that
these functions behave more like exponential functions which shows
that the coupling between the BD field and matter increases with
time. In figures 7 and 8, we provide the behavior of the BD
potential which is increasing (decreasing) for $V_+$ ($V_-$) against
time for large (small) values of the BD parameter. Figure 9 gives
the variation of the BD potential against the BD field while the
figure 10 provides the variation of the scale factor in the limit
when the BD parameters vanishes (becomes negligible).

In figures 11, 12 and 13, we have plotted statefinder parameters and
the deceleration parameter of our model against the BD parameter. It
is observed that deceleration parameter is positive, thereby
producing a decelerated Universe. Though the model suggests the
expansion of the Universe, yet the model does not explain cosmic
acceleration.  The stability analysis of the solutions is important.
But it will be beyond the scope of our paper. Such analysis can be
done using the similar methodology, as it has been done in
\cite{stability1}, by defining dimensionless variables. The
stability of BD with Chameleon field has been done already in
\cite{stability2} for the same system of equations as we have.
\section*{Acknowledgement}
The authors would like to thank anonymous referees  for helpful
comments and suggestions.

\end{document}